\documentclass[aps,amsmath,amssymb,amsfonts,preprint,superscriptaddress,citeautoscript,byrevtex, onecolumn]{revtex4-1}
\usepackage{graphicx,color}
\usepackage{epstopdf}
\pdfoutput=1 
\usepackage{soul}
\usepackage{ulem}

\begin{document}
\title{ IrCrMnZ (Z=Al, Ga, Si, Ge) Heusler alloys as electrode materials for MgO-based magnetic tunneling junctions:\\ A first-principles study}
\author{Tufan Roy}
\email{roy.tufan.a3@tohoku.ac.jp}
\affiliation{Research Institute of Electrical Communication (RIEC), Tohoku University, Sendai 980-8577, Japan}
\author{Masahito Tsujikawa}
\affiliation{Research Institute of Electrical Communication (RIEC), Tohoku University, Sendai 980-8577, Japan}
\affiliation{Center for Spintronics Research Network (CSRN), Tohoku University, Sendai 980-8577, Japan}
\author{Masafumi Shirai}
\affiliation{Research Institute of Electrical Communication (RIEC), Tohoku University, Sendai 980-8577, Japan}
\affiliation{Center for Spintronics Research Network (CSRN), Tohoku University, Sendai 980-8577, Japan}
\affiliation{Center for Science and Innovation in Spintronics (CSIS), Core Research Cluster (CRC), Tohoku University, Sendai 980-8577, Japan}
\begin{abstract}
We study IrCrMnZ (Z=Al, Ga, Si, Ge) systems using first-principles calculations from the perspective of their application as the electrode materials of MgO-based MTJs. These materials have highly spin-polarized conduction electrons with partially occupied $\Delta_1$ band, which is important for coherent tunneling in parallel magnetization configuration. The Curie temperatures of IrCrMnAl and IrCrMnGa are very high (above 1300 K) as predicted from mean-field-approximation. The stability of ordered phase against various antisite disorders has been investigated. We discuss here the effect of ``spin-orbit-coupling'' on the electronic structure around Fermi level. Further, we investigate the electronic structure of IrCrMnZ/MgO heterojunction along (001) direction. IrCrMnAl/MgO and IrCrMnGa/MgO maintain half-metallicity even at the MgO interface, with no interfacial states at/around Fermi level in the minority-spin channel. Large majority-spin conductance of IrCrMnAl/MgO/IrCrMnAl and IrCrMnGa/MgO/IrCrMnGa is reported from the calculation of ballistic spin-transport property for parallel magnetization configuration. We propose IrCrMnAl/MgO/IrCrMnAl and IrCrMnGa/MgO/IrCrMnGa as promising MTJs with a weaker temperature dependence of tunneling magnetoresistance ratio, owing to their very high Curie temperatures.
\end{abstract}


\maketitle

\section{Introduction} 
In the field of spintronics, the family of Heusler alloy has been extensively studied, since last few decades. Co-based Heusler alloys have attracted prime attention of the researchers, because of their suitability in terms of highly spin-polarized conduction electrons, and high Curie temperature ($T_\mathrm {C}$), which are the crucial prerequisites for spintronics application to obtain spin-polarized current at room temperature.\cite{Miura-PRB-2011,Miura-JPCM-2009,Miura-prb,PRL-Hulsen,JAP-Ishikawa-2008,APL-Liu-2012,PRM-Tsuchiya} The most studied Heusler alloy as the electrode material of MgO-based magnetic tunnel junctions (MTJs) is Co$_2$MnSi.\cite{Miura-PRB-2011,Saito-PRB-2010} It is a long-standing challenge for the researchers to obtain large tunneling magnetoresistance (TMR) ratio at room temperature. Till date the highest TMR ratio has been reported is about 2600\% at low temperature using Co$_2$MnSi or that partially substituted by Fe for Mn, as electrode, and MgO as a barrier material.\cite{Liu-JPD-2015} However, the TMR ratio falls rapidly with an increase of temperature, which acts as a hindrance for their practical application.\cite{Hu-PRB-2016}

In literature there have been several arguments to address the source of reduction of TMR ratio. One possible reason is that the highly spin-polarized electrode looses its half-metallic nature at the interface of its heterojunction with MgO. It is claimed that the interfacial states in the minority-spin channel could act as a spin-flip center at finite temperature and degrades the spin-filtering property of the MTJs.\cite{Mavropoulos-PRB-2005} Therefore, along with half-metallic electrode, it is also important to preserve the half-metallicity even at the interface for a better performance of MTJs.

Misfit dislocation resulting from lattice mismatch between electrode and barrier material, is also regarded as another source of reduction of TMR ratio.\cite{Bonell-IEEE, Bonell-PRB,Kunimatsu-apex} Lattice mismatch results in a misfit dislocation at the interface, and could affect coherent tunneling through the barrier. This lattice mismatch is significantly large in case of Co$_2$MnSi/MgO (001) MTJ, which is about 5\%. Recently, we showed from the first-principles calculations that replacing one Co atom by Ir atom of well-known Co$_2$-based Heusler alloys could effectively reduce the interfacial lattice mismatch.\cite{troy-jmmm-2020} Later on, we successfully fabricated CoIrMnAl thin film. However, the Curie temperature ($T_\mathrm {C}$) for CoIrMnAl is found to be somewhat in the moderate range ($\approx$ 400 K).\cite{Momma-jalcom-2021}

Recently, there are significant amount of studies on the ferrimagnetic Heusler alloys both from the experiment and first-principles calculations from the perspective of the spintronics application.\cite{Betto-aip-adv-2016,kudo-prb-2021,onodera-jjap,troy-jpcm,umetsu-jpcm} 
Ferrimagnetic systems are of special interest in the spin transfer toque magnetoresistive memories, because of their low magnetization, which require low magnetization switching current.\cite{balke-apl, mizukami-prb, mizukami-prl}

In this study, we focused on the ferrimagnetic Heusler alloys IrCrMnZ (Z= Al, Ga, Si, and Ge), which have excellent lattice matching with MgO. First we confirm the stability of this family in terms of the formation energy and phase separation energy. Dynamical and mechanical stability of these systems are confirmed by phonon dispersion and elastic constants. After confirming the stability of these ferrimagnetic Heusler alloys with very high $T_\mathrm {C}$, we discuss the electronic structure of IrCrMnZ/MgO heterojunctions. Finally, after calculating spin-dependent ballistic transport properties, we propose IrCrMnZ/MgO/IrCrMnZ (001) MTJs to be promising candidates in spintronics applications.

\section{Method}

Structural optimization of the bulk phase has been performed using Vienna \textit{ab-initio}
Simulation Package 
(VASP)\cite{VASP} in combination with projector augmented wave (PAW) 
method.\cite{PAW} A generalized gradient approximation (GGA) has been exploited for the exchange correlation potential/energy.\cite{PBE}
 We have used an energy cut-off of 
500\,eV for the planewaves. The final energies have 
been calculated with a $k$-mesh of 
16$\times$16$\times$16 for the 
bulk phase. The 
energy and the force tolerance for our calculations 
were 10 $\mu$eV and 10 meV/\AA, respectively.

Phonon calculations were carried out within small displacement method using PHONOPY code\cite{phonopy}, in combination with VASP. We used a 2$\times$2$\times$2 supercell for phonon calculations. 
Furthermore to investigate the effect of swapping (antisite) disorder on the properties of IrCrMnZ alloys, we adopted supercell approach. Supercells contain 64 atoms. Our considered supercells are special quasi-random structures (SQS) to model the chemical disorder.\cite{SQS} We used Alloy Theory Automated Toolkit (ATAT) package to generate the SQSs.\cite{atat}

We construct a heterojunction of IrCrMnZ and MgO along (001). 
Here, we consider eleven layers of Heusler alloys and five layers of MgO, as considered in Ref \onlinecite{troy-jpcm}. 
For the structural optimization of the junction along $c$-axis we use a $k$-mesh of 10$\times$10$\times$1. A denser $k$-mesh 16$\times$16$\times$2 has been used for the further calculations of electronic structure and magnetic properties of the heterojunction.

We used Green's function based full potential spin-polarized relativistic Korringa-Kohn-Rostoker method as implemented in the SPR-KKR programme package,\cite{sprkkr} for the evaluation of Heisenberg exchange coupling constants between the atoms of bulk IrCrMnZ (Z=Al, Ga, Si, Ge). Following Liechtenstein's approach,\cite{licechenstein} we evaluate $T_\mathrm {C}$ within mean-field-approximation (MFA). We used 917 $k$-points for the Brillouin zone integration. For angular momentum cut-off $l_{max}$=3 was considered for each atom. We used GGA exchange correlation potential/energy.\cite{PBE} For the determination of Fermi energy ($E_\mathrm {F}$), we employed Llyod's formula.\cite{lloyd1,lloyd2}

Furthermore, the first-principle calculations of ballistic conductance has been carried out using PWCOND code \cite{PWCOND} in QUANTUM ESPRESSO package\cite{QE}, following the method suggested by Choi and Ihm\cite{Choi-PRB}.
The GGA exchange correlation potential is also used here.\cite{PBE} The cut-off energy for the wave function and charge density are set to 50 Ry and 500 Ry, respectively. The detail of the calculational method for ballistic conductannce could be found in References \onlinecite{Miura-JPCM-2009} and \onlinecite{Miura-prb}.

\section{Results and Discussion}

\subsection{Bulk phase}
\subsubsection{Crystal structure and stability analysis} In the ordered phase, quaternary Heusler alloys IrCrMnZ (Z=Al, Ga, Si, Ge) possess LiMgPdSn-type of crystal structure with F$\bar{4}$3m space group. These structures have four interpenetrating fcc sublattices with their origins at, $A$ (0, 0, 0), $B$ (0.5, 0.5, 0.5), $C$ (0.25, 0.25, 0.25), and $D$ (0.75, 0.75, 0.75), respectively. To obtain the ground state crystal structure, we considered three different configurations, where $A$, $B$, and $C$ sites are occupied by (a) Configuration I: Ir, Cr, and Mn atoms, (b) Configuration II: Ir, Mn, and Cr atoms, and (c) Configuration III: Mn, Cr, and Ir atoms, respectively. In all the cases Z atom occupies the $D$ site. In Table 1, we summarize the formation energies of the respective Heusler formula unit (X{X$^\prime$}YZ), for all three configurations. Formation energy ($\Delta E$) is evaluated from the following formula:

\begin{table*}[hbt!]
\renewcommand{\thetable}{\arabic{table}}

\centering
\caption{ Formation energies of IrCrMnZ Heusler alloys, for different configurations, in the unit of eV/f.u.}
\begin{tabular}{|c|c|c|c|c|c|c|c|c|}
\hline Material & Configuration I   & Configuration II & Configuration III\\
\hline IrCrMnAl & -1.558 & -1.288  & +0.067 \\
\hline IrCrMnGa &-1.065 & -0.654& + 0.098\\
\hline IrCrMnSi &-1.551 &-1.263  & -0.023 \\
\hline IrCrMnGe &-0.762 &-0.388  &+0.634 \\

\hline
\end{tabular}   
\end{table*}
\begin{equation}
\Delta E = E^{{X{X}^\prime}YZ}-E^{X}-E^{X^\prime}-E^{Y}-E^{Z}
\end{equation}

Here, $E^{X{X^\prime}YZ}$, $E^{X}$, $E^{X^\prime}$, $E^{Y}$, $E^{Z}$ are the total energies of the X{X$^\prime$}YZ Heusler alloy, and each constituent element, i.e., X, X$^\prime$, Y, and Z in the respective ground state bulk phase.
From the definition of formation energy, it is evident that more negative value of formation energy corresponds to higher stability of the particular configuration. For all of the IrCrMnZ systems, we find that Configuration I is the most stable one. Hereafter, we focus our attention on this configuration.

Note that the formation energy does not take care of possible decomposition of the IrCrMnZ alloys into other alloys/compounds. Recently, we reported phase separation of computationally predicted Heusler alloy, NiCrMnSi, into other alloys/compounds during its experimental fabrication, which was supported by positive phase separation energy.\cite{onodera-jjap}
 Therefore, to discuss phase stability of IrCrMnZ alloys, we further calculate phase separation energy ($\delta E$). $\delta E$ compares the stability of IrCrMnZ alloys with respect to other stable alloys composed of Ir, Mn, Cr, or Z elements. For the sake of simplicity, we discuss the stability of IrCrMnZ systems against the decomposition into binary alloys, namely, IrMn, Al$_3$Cr, Ga$_4$Cr, Cr$_3$Si, Cr$_3$Ge, depending on the Z. The structural parameters of these binary alloys are adopted from the Materials Project database.\cite{mp} We used the following equations to evaluate phase separation energy ($\delta E$):
 
\begin{equation}
\delta E_{\rm IrCrMnAl}= \Delta E_{\rm IrCrMnAl} - \left[\Delta E_{\rm MnIr} +\dfrac{1}{3}\Delta E_{\rm Al_{3}Cr} +\dfrac{2}{3}\Delta E_{\rm Cr}\right]
\end{equation}
\begin{equation}
\delta E_{\rm IrCrMnGa}= \Delta E_{\rm IrCrMnGa} - \left[\Delta E_{\rm MnIr} +\dfrac{1}{4}\Delta E_{\rm Ga_{4}Cr} +\dfrac{3}{4}\Delta E_{\rm Cr}\right]
\end{equation}
\begin{equation}
\delta E_{\rm IrCrMnSi}= \Delta E_{\rm IrCrMnSi} - \left[\Delta E_{\rm MnIr} +\dfrac{1}{3}\Delta E_{\rm Cr_{3}Si} +\dfrac{2}{3}\Delta E_{\rm Si}\right]
\end{equation}
  
\begin{equation}
\delta E_{\rm IrCrMnGe}= \Delta E_{\rm IrCrMnGe} - \left[\Delta E_{\rm MnIr} +\dfrac{1}{3}\Delta E_{\rm Cr_{3}Ge} +\dfrac{2}{3}\Delta E_{\rm Ge}\right]
\end{equation}

Here, $\Delta E_{\rm MnIr}$, $\Delta E_{\rm Al_{3}Cr}$, $\Delta E_{\rm Ga_{4}Cr}$, $\Delta E_{\rm Cr_{3}Si}$, and $\Delta E_{\rm Cr_{3}Ge}$ are the formation energies for MnIr, Al$_3$Cr, Ga$_4$Cr, Cr$_3$Si, and Cr$_3$Ge, respectively, which are found to be $-0.272$ eV/f.u., $-0.578$ eV/f.u., $-0.848$ eV/f.u., $-1.439$ eV/f.u., $-0.510$ eV/f.u. Note that $\Delta E_{\rm Cr}$, $\Delta E_{\rm Si}$, $\Delta E_{\rm Ge}$ are the formation energies of Cr, Si, and Ga which are 0 eV/f.u. The formation energies of IrCrMnZ alloys are adopted from Table 1, corresponding to Configuration I. After substituting these values in the above equations, we obtain $\delta E$, for IrCrMnAl, IrCrMnGa, IrCrMnSi, and IrCrMnGe as $-1.093$ eV/f.u., $-0.581$ eV/f.u., $-0.799$ eV/f.u., and $-0.320$ eV/f.u. Negative values of $\delta E$ signify that IrCrMnZ alloys are not likely to decompose into the considered binary alloys upon their experimental fabrication.

\begin{table*}[hbt!]
\renewcommand{\thetable}{\arabic{table}}

\centering
\caption{ Elastic constants of IrCrMnZ Heusler alloys.}
\begin{tabular}{|c|c|c|c|c|c|c|c|c|c|c|}
\hline Material &  $C_{11}$  &   $C_{12}$&   $C_{44}$&Bulk modulus& Shear modulus\\
&(GPa)&(GPa)&(GPa)&(GPa)&(GPa)\\
\hline IrCrMnAl & 215 & 141  & 133 &166&95\\
\hline IrCrMnGa &204 & 150& 125&168&86\\
\hline IrCrMnSi &274 &178  & 145&210&106 \\
\hline IrCrMnGe &255 &152  &122&185&94 \\

\hline
\end{tabular}   
\end{table*}

\begin{figure}[h]
\includegraphics[width=0.7\textwidth]{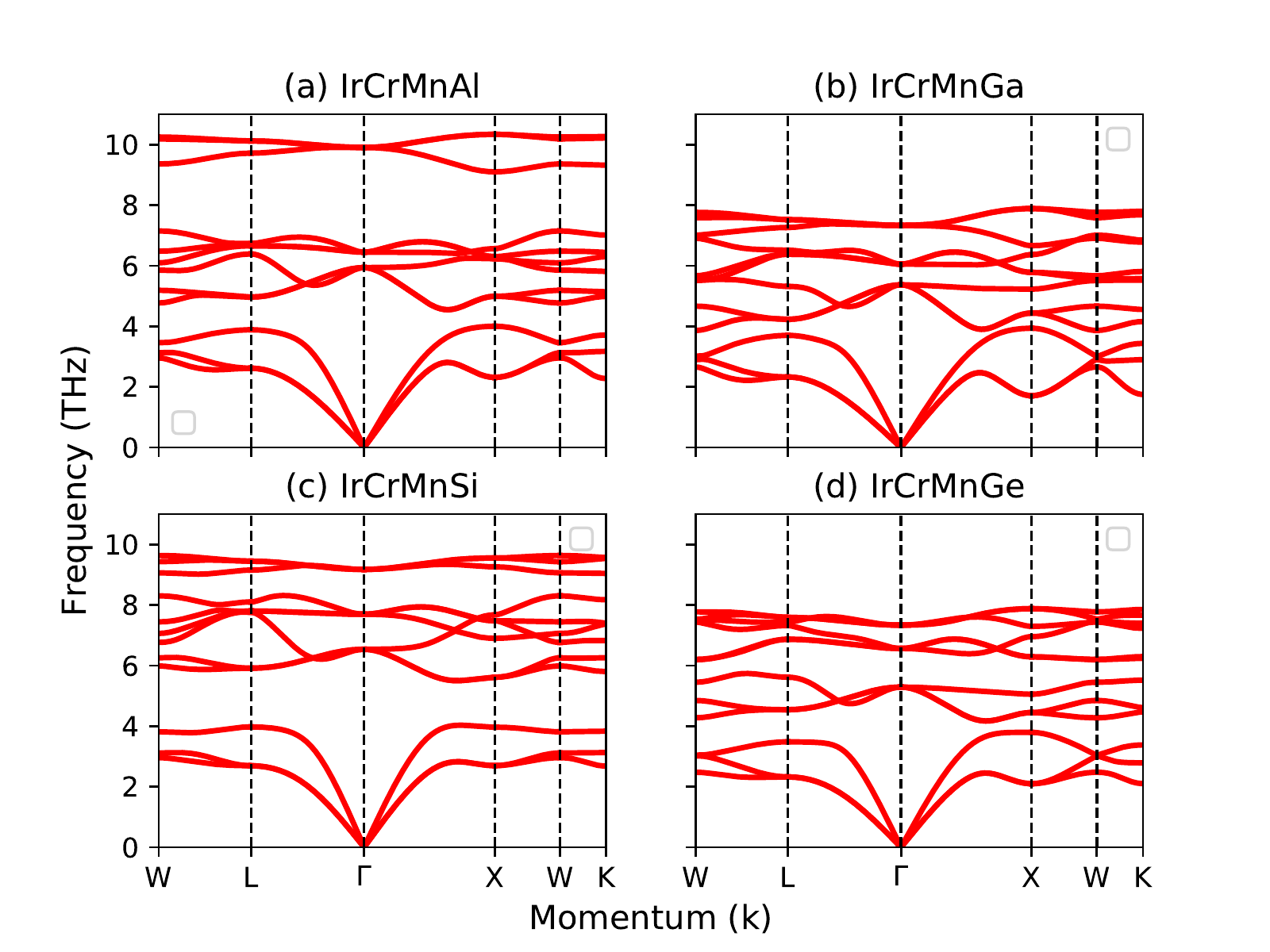}
\caption
{(Color online) Phonon dispersion curves for (a) IrCrMnAl, (b) IrCrMnGa, (c) IrCrMnSi, and (d) IrCrMnGe, respectively. } 

\end{figure}

Now, we discuss the mechanical stability of these systems based on elastic constants and phonon dispersion curves. As, we previously mentioned that IrCrMnZ systems have cubic ground state. So, there are three independent elastic constants related to these systems, namely $C_{11}$, $C_{12}$, and $C_{44}$. The criteria for mechanical stability of a cubic solid is given by $C_{11} > 0$, $C_{44} > 0$, $C_{11} - C_{12} > 0$, and $C_{11} + 2C_{12} > 0$. The method of calculation has been discussed in detail in Reference \onlinecite{troy-prb-2016}.

From Table 2 , it can be clearly observed that the criteria for mechanical stability are satisfied by all these systems. As these systems have been investigated for the first time, so there are no references to compare these values. It is highly desired to fabricate and measure the elastic properties of the IrCrMnZ systems.

The dynamical stability of the system are further discussed in terms of the phonon dispersion curves as presented in figure 1. It is to be noted that there are total twelve number of branches owing to four atoms in the primitive unit cell of the quaternary Heusler alloys. The phonon dispersion curves are plotted along the high symmetry directions. We find that all phonon modes have the positive frequencies, suggesting the dynamical stability of the IrCrMnZ systems.

\subsubsection{Magnetic properties}
After confirming the mechanical stability, here, we discuss about the magnetic properties. In Table 3, we summarize, the lattice parameter ($a$), the total magnetic moments ($\mu_{total}$) and that of each constituent element, $T_\mathrm {C}$, and spin polarization ($P$) at $E_\mathrm {F}$. It is to be noted that all these systems have lattice parameters in the range of 5.88 \AA{} to 6.03 \AA{}, which are significantly large compared to Co$_2$MnSi (5.65 \AA). It could be helpful for better compatibility of these systems to form heterojunction with MgO. It is to be mentioned that there is perfect lattice matching between Heusler alloy and MgO heterojunction along (001) direction, if the former one has lattice parameter of 5.95 \AA. 
First-principles study by K\"ubler \textit{et al.} suggests that in case of X$_2$MnZ systems, X atom predominantly determines the lattice parameter of the system.\cite{prb-kubler-1983}
From our previous study as well the present work it can be conjectured that in case of quaterneray Heusler alloy (XX'YZ), either X or X' should be 5$d$ element, to minimize the lattice mismatch with MgO. However, the replacement of one Co of traditional Co$_2$-based Heusler alloy by Ir reduces the $T_\mathrm {C}$ of the electrode as we observed in case of CoIrMnAl.\cite{troy-jmmm-2020,Momma-jalcom-2021} Note that $T_\mathrm {C}$ is one of the most important parameters to control the temperature dependence of TMR ratio of a MTJ. So, along with minimizing the lattice mismatch at the MgO interface, by choosing 5$d$ element (e.g. Ir) as X or X', one must find an alternative route to compensate the reduction of $T_\mathrm {C}$, resulting from this substitution. In the following section we will discuss the $T_\mathrm {C}$ of IrCrMnZ system.

\begin{table*}[hbt!]
\renewcommand{\thetable}{\arabic{table}}

\centering
\caption{ Calculated bulk properties of IrCrMnZ}
\begin{tabular}{|c|c|c|c|c|c|c|c|c|}
\hline Material & $a$  & $\mu_{total}$ & $\mu_{\mathrm Ir}$ & $\mu_{\mathrm Cr}$ & $\mu_\mathrm {Mn}$ & $\mu_{Z}$& $T_\mathrm C$ & $P$ (\%)\\
&(\AA)&($\mu_{\mathrm B}$)&($\mu_{\mathrm B}$)&($\mu_{\mathrm B}$)&($\mu_{\mathrm B}$)&($\mu_{\mathrm B}$)&(K)&\\
\hline IrCrMnAl & 6.008 &1.01 &0.20& -2.42 & 3.20 & 0.01 &1338&95\\
\hline IrCrMnGa & 6.029&1.04 & 0.21& -2.49& 3.26 & 0.01 &1355&90\\
\hline IrCrMnSi & 5.879 &2.00 & 0.20&-1.30 & 2.98 & 0.03 &826& 100\\
\hline IrCrMnGe & 5.993 &2.00 & 0.24& -1.57 & 3.18 & 0.04 &1019&100\\

\hline
\end{tabular}   
\end{table*}

Note that the $\mu_{total}$ of half-metallic full-Heusler alloys obey Slater-Pauling rule,\cite{Galanakis-prb-2002} according to which, $\mu_{total}$ is related to the number of valence electrons ($N_v$) by the following relation: $\mu_{total}$ = $N_v$ -24. In case of IrCrMnGa and IrCrMnAl, $N_v$ =25, and for rest of the two systems $N_v$ =26. Note that, $\mu_{total}$ is very closed to 1.00 $\mu_{\mathrm B}$ for IrCrMnGa and IrCrMnAl. $\mu_{total}$ is exactly 2.00 $\mu_{\mathrm B}$ for IrCrMnSi and IrCrMnGe, owing to their 100\% spin polarization. It is to be noted for all these cases Cr atom has an anti-parallel spin alignment with respect to Mn and Ir atoms. This leads to a ferrimagnetic ground state for all the cases. In case of Mn-based Heusler alloys, it has been shown that if Mn has octahedral coordination with respect to the main group element (i.e. Z atom), the magnetic moment of Mn is largely localized and for tetrahedrally coordinated case it has itinerant type of magnetism.\cite{karel-pccp-2017}
For the IrCrMnZ systems, Mn has octahedral coordination and on the other hand Cr has tetrhedral coordination with respect to Z atoms. It results in localized moment of Mn atom and the magnetic moment is kind of robust against the Z atom. In contrast the magnetic moment of Cr atom is sensitive to the number of valence electrons of Z atom. Note that, in case of IrCrMnSi, the magnetic moment of Mn atom is slightly less than the other cases, which is because of smaller lattice parameter, and results in less localization of moment.

We evaluate $T_\mathrm {C}$ within mean-field-approximation (MFA) following Liechtenstein formalism.\cite{licechenstein}
For the electrode material of a MTJ, it is necessary that $T_\mathrm {C}$ should be very high as it minimizes the thermal fluctuation of the magnetic moments and improves the temperature dependence of TMR ratio. The temperature dependence of TMR ratio of bcc Co/MgO/bcc Co is reported to be very small, owing to its very high $T_\mathrm {C}$ ($\approx$ 1500 K from MFA).\cite{Yuasa-apl-2006, Lezaic-apl-2007} Among the Heusler alloys, Co$_2$MnSi is the most studied as electrode of MgO-based MTJs. Experimental $T_\mathrm {C}$ of Co$_2$MnSi is reported as 985 K, although MFA shows the $T_\mathrm {C}$ is about 1170 K.\cite{chen-sci-rep-2018,webster-jpcs}
In case of IrCrMnZ Heusler alloys, $T_\mathrm {C}$ are very high. In case of IrCrMnAl and IrCrMnGa, $T_\mathrm {C}$ are 1338 K and 1355 K, which are even higher than Co$_2$MnSi. For IrCrMnSi and IrCrMnGe, the $T_\mathrm {C}$ are 826 K and 1019 K, respectively. A strong antiferromagnetic exchange coupling between nearest neighboring Mn and Cr spins, results in high $T_\mathrm {C}$ for these systems. Note that, $T_\mathrm {C}$ is the lowest for IrCrMnSi. In this case, smaller absolute values of Cr and Mn magnetic moments results in weaker exchange energy compared to rest of the systems, hence the lowest $T_\mathrm {C}$. Balke \textit{et al.} have shown that $T_\mathrm {C}$ increases with increase of $\mu_{total}$ for several ferromagnetic Co-based Heusler alloys.\cite{Balke-Sci-Tech-2008} Co$_2$FeSi is reported to have highest $\mu_{total}$ = 6 $\mu_{\mathrm B}$, and its experimental $T_\mathrm {C}$ is about 1100 K.\cite{apl-sabine-2006} In our study, the magnetic ground state of IrCrMnZ systems are ferrimagnetic, which results in a lower $\mu_{total}$. Ferromagnets and/or ferrimagnets with low net magnetic moment ($\mu_{total}$) are desirable for electrode materials of MTJs, since those give rise to lower stray fields as well as lower threshold fields required for magnetization switching by spin-transfer torques. Both IrCrMnAl and IrCrMnGa fulfill this criterion as their $\mu_{total}$ is as small as 1 $\mu_{\mathrm B}$. On the other hand $T_\mathrm {C}$ for IrCrMnAl and IrCrMnGa are higher than 1300 K. This is advantageous for these two systems for application to the electrodes of MTJs.

\begin{figure}[h]

\includegraphics[width=0.7\textwidth]{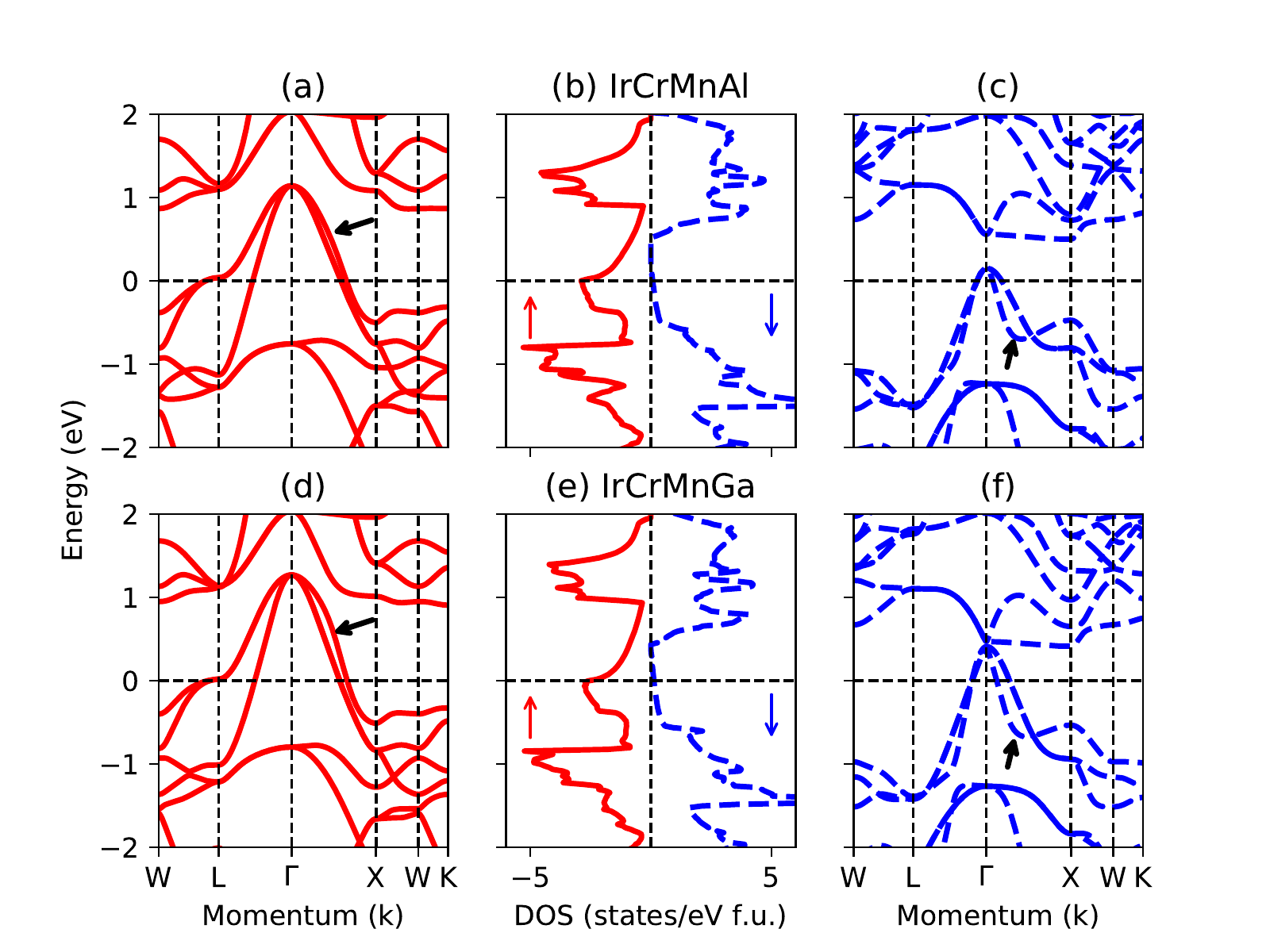}
\caption
{(Color online) Upper panel: Band dispersion curves for (a) majority-spin states, (c) minority-spin states, and (b) density of states of IrCrMnAl. 
Lower panel: Band dispersion curves for (d) majority-spin states, (f) minority-spin states, and (e) density of states of IrCrMnGa.
The Fermi level is at 0 eV.} 

\end{figure}

\begin{figure}[h]

\includegraphics[width=0.7\textwidth]{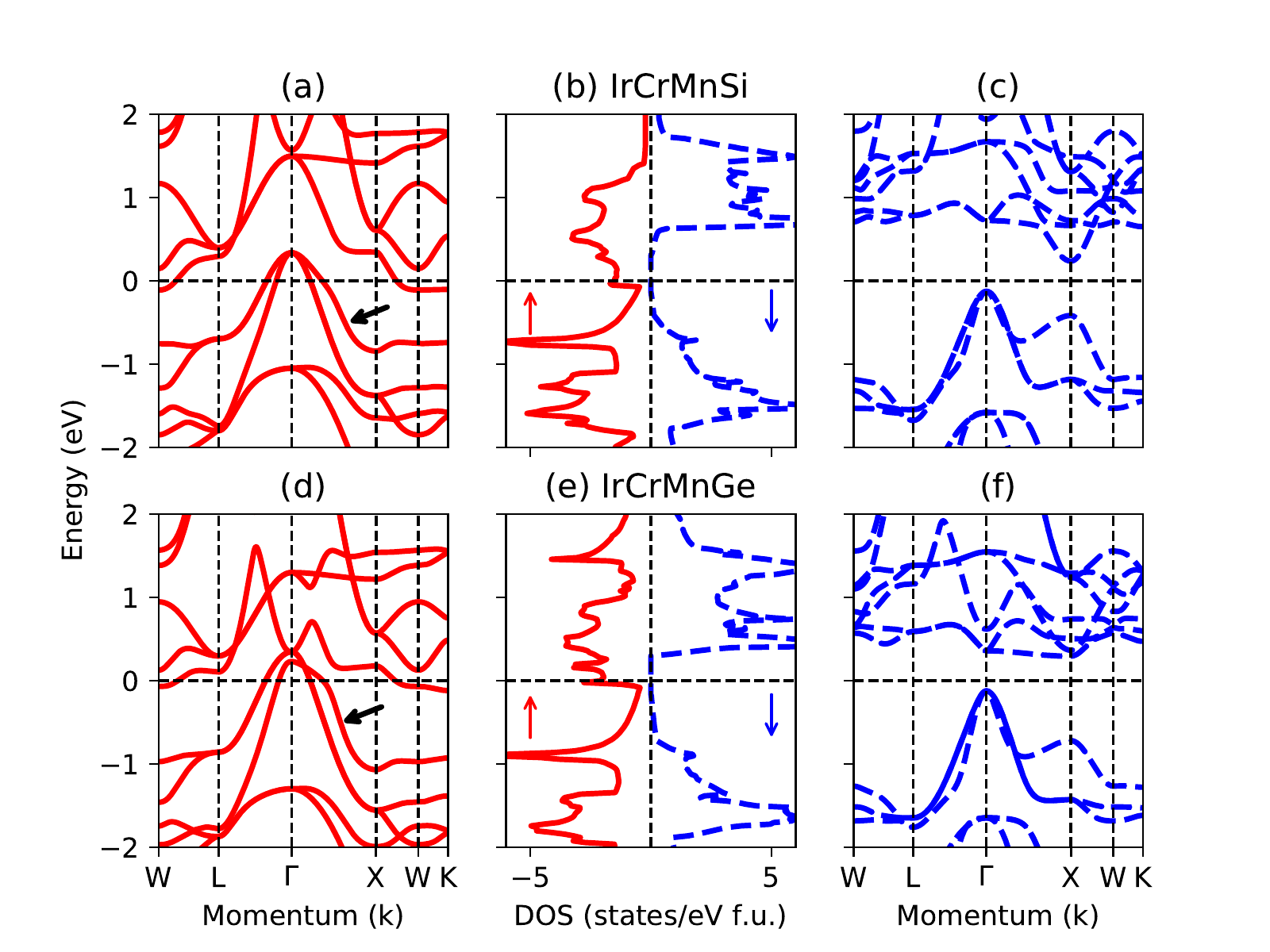}
\caption
{(Color online) Upper panel: Band dispersion curves for (a) majority-spin states, (c) minority-spin states, and (b) density of states of IrCrMnSi. 
Lower panel: Band dispersion curves for (d) majority-spin states, (f) minority-spin states, and (e) density of states of IrCrMnGe.
The Fermi level is at 0 eV.} 
\end{figure}

\subsubsection{Electronic structure}

 Figure 2 depicts the spin-resolved electronic band structure and density of states (DOS) for IrCrMnAl and IrCrMnGa systems. Figure 3 shows the same for IrCrMnSi and IrCrMnGe systems. 
 From figure 2, it is obvious that both IrCrMnAl and IrCrMnGa show metallic band structure for the majority-spin channel. We find that partially filled band crosses the Fermi level along $\Gamma$-L and $\Gamma$-X directions. For the minority-spin channel the top of the valence band just touches $E_\mathrm {F}$ at $\Gamma$ point leading to a negligibly small DOS (0.07 states/ev f.u) compared to the value of metallic majority-spin channel (2.89 states/eV f.u.) at $E_\mathrm {F}$. It leads to a very high spin polarization of about 95\% at $E_\mathrm {F}$. In case of IrCrMnGa, the electronic structure is very similar to its isoelectronic IrCrMnAl system. Although the minority-spin DOS at $E_\mathrm {F}$ increases a bit (0.13 states/eV f.u.), it is still very small compared to its value for majority-spin channel (2.64 states/eV f.u.), leading to relatively high spin polarization (90 \%) at $E_\mathrm {F}$.

 In figure 3, we present the electronic structure of IrCrMnSi and IrCrMnGe systems. Similar to the IrCrMnAl and IrCrMnGa, the majority-spin channel is metallic for IrCrMnSi and IrCrMnGe. However, for both of these systems there are no electronic bands in the minority-spin channel, crossing $E_\mathrm {F}$, which results in a 100\% spin polarization.

As mentioned before, the motivation of this work is to assess the feasibility of these Heusler alloys as electrodes in MgO-based MTJs along the (001) direction. Thus, the electronic structure along the $\Gamma$-X ($\Delta$ line) is of special interest. We identify the presence of electronic bands with $\Delta_1$ symmetry (i.e. $s$, $p_z$, $d_{z^2}$) by the black solid arrow for all the cases in figure 2 and figure 3. In literature, it is well established that in case of crystalline MgO barrier, presence of spin-polarized $\Delta_1$ band is very important to obtain large TMR ratio.\cite{butler,parkin-nature,apl-Djayaprawira} Note that in case of IrCrMnGa and IrCrMnAl there is a highly spin-polarized $\Delta_1$ band, and in case of IrCrMnSi and IrCrMnGa $\Delta_1$ band is fully spin-polarized. We will discuss the impact of the presence of the highly spin-polarized $\Delta_1$ band on the spin-dependent transport properties in the following part of the manuscript.
    \begin{figure}[h]
\includegraphics[width=1.0\textwidth]{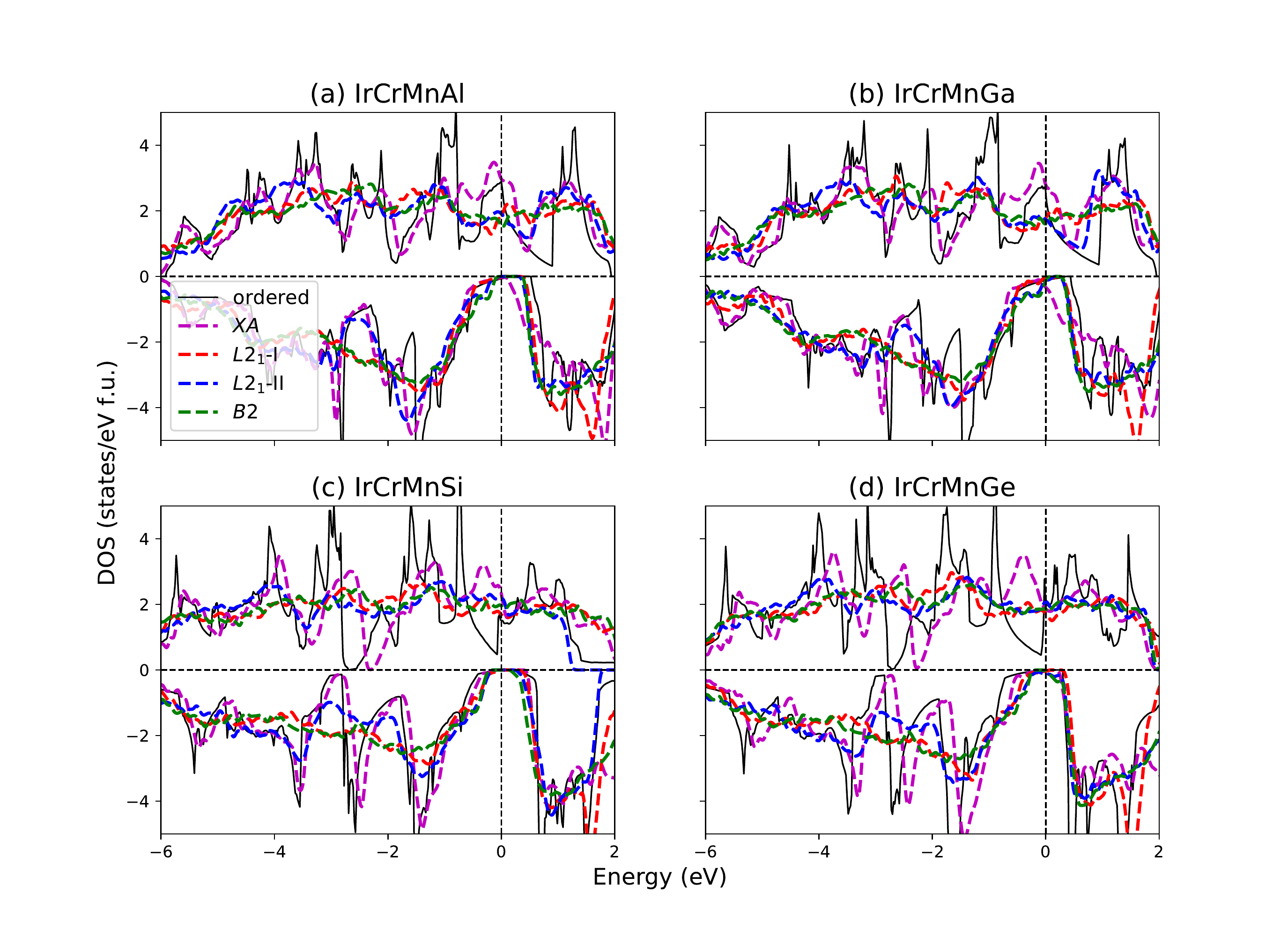}
\caption
{(Color online) Comparison of spin-resolved total density of states of ordered phase, Cr-Mn disorder ($XA$), Ir-Cr disorder ($L$2$_1$-I), Mn-Z disorder ($L$2$_1$-II), Ir-Cr and Mn-Z disorder simultaneously ($B$2) of (a) IrCrMnAl, (b) IrCrMnGa, (c) IrCrMnSi, and (d) IrCrMnGe, respectively.} 

\end{figure}
 
 \begin{figure}[h]
\includegraphics[width=0.8\textwidth]{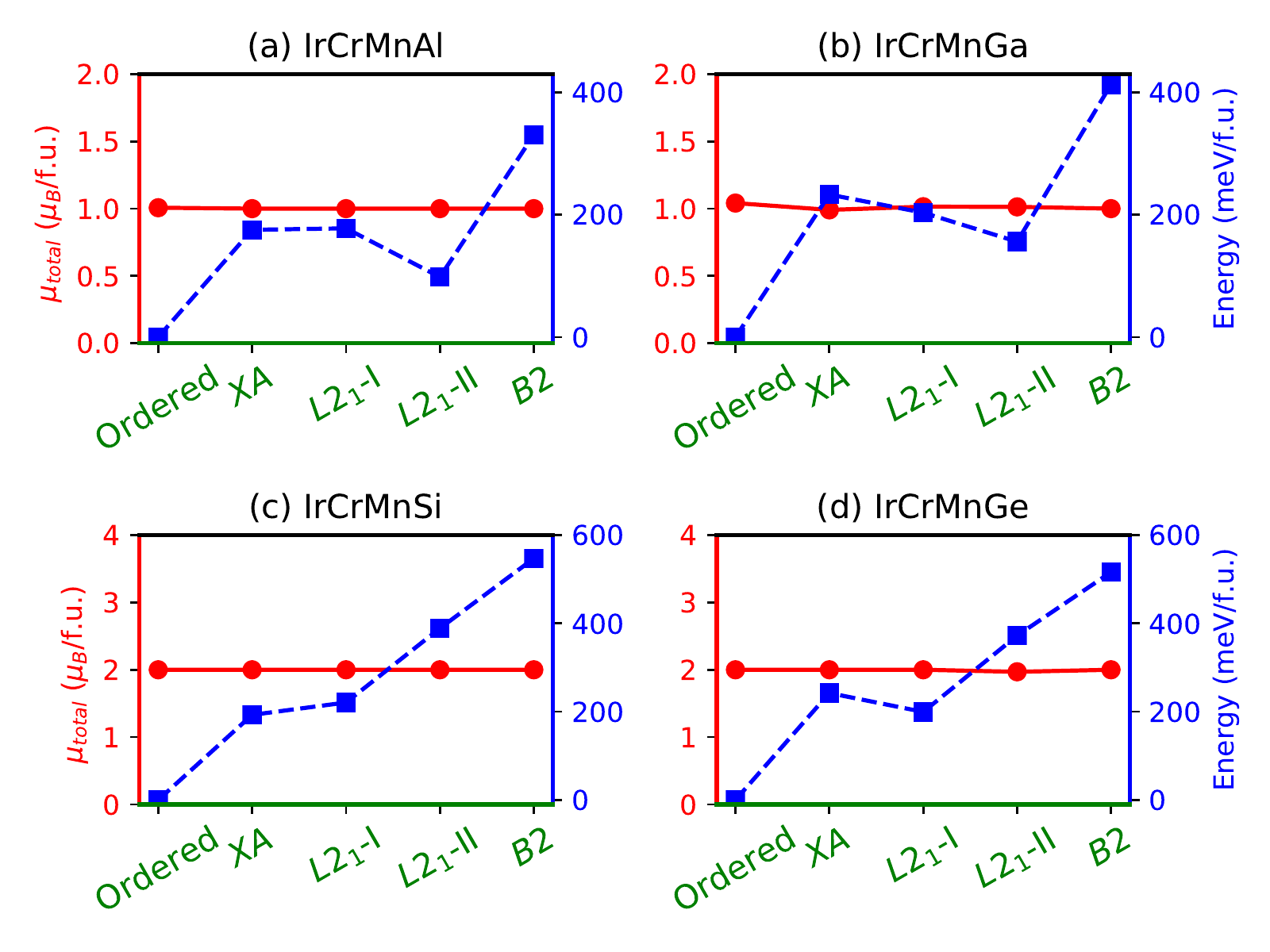}
\caption
{(Color online) Comparison of total magnetic moment and total energy of Cr-Mn disorder ($XA$), Ir-Cr disorder ($L$2$_1$-I), Mn-Z disorder ($L$2$_1$-II), Ir-Cr and Mn-Z disorder simultaneously ($B$2) with respect to ordered phases ($Y$) for (a) IrCrMnAl, (b) IrCrMnGa, (c) IrCrMnSi, and (d) IrCrMnGe, respectively.} 

\end{figure}
 \begin{figure}[h]
\includegraphics[width=0.7\textwidth]{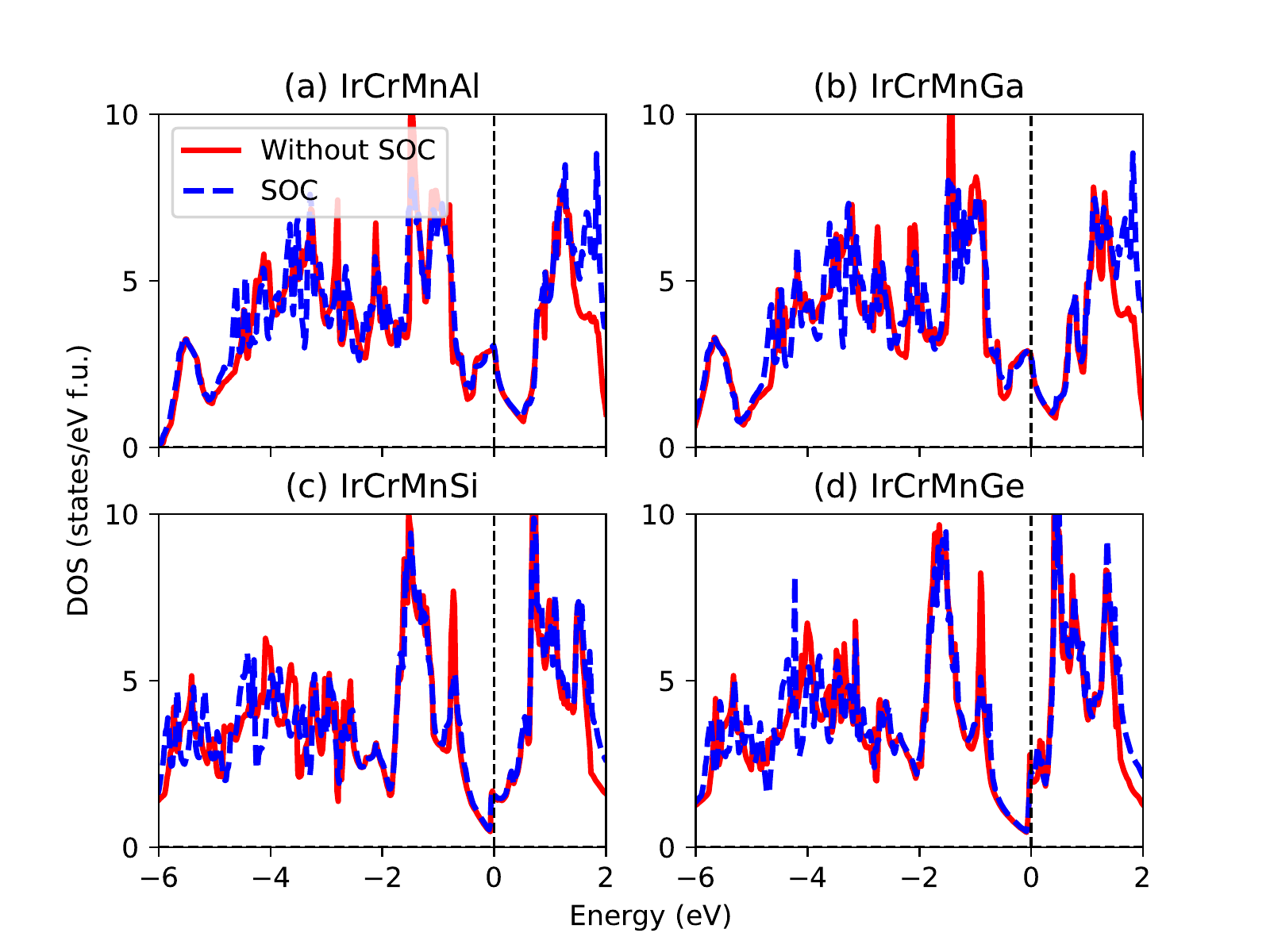}
\caption
{(Color online) Comparison of total density of states with and without the effects of ``spin-orbit-coupling'' for (a) IrCrMnAl, (b) IrCrMnGa, (c) IrCrMnSi, and (d) IrCrMnGe, respectively. The Fermi level is at 0 eV.} 

\end{figure}
 
 It is also important to discuss the effects of various swapping disorder on the electronic structure. Here, we considered four different types of disorders. The considered disordered cases are (a) full-swap disorder between Cr and Mn atoms; (b) full-swap disorder between Ir and Cr atoms; (c) full-swap disorder between Mn and Z atoms; (d) full-swap disorder between Ir and Cr atoms, and between Mn and Z atoms, simultaneously. Note that the full-swap between Cr and Mn atoms transforms ordered $Y$ structure into $XA$ structure. In both the cases of Ir-Cr and Mn-Z swapping disorders, the ordered $Y$ structure is transformed into $L$2$_1$ structure.
 We name structure after swapping disorder between Ir and Cr atoms as $L$2$_1$-I structure, on the other hand, the structure resulting from swapping disorder between Mn and Z atom is named as $L$2$_1$-II structure. In the case of appearance of both Ir-Cr and Mn-Z disorders, simultaneously, the system is reduced to a $B$2 structure.

 Figure 4 shows the impact of the various swapping disorders on the electronic structure. It could be noticed that for the considered swapping disorders i.e., $XA$, $L$2$_1$-I and $L$2$_1$-II structures, the width of the half-metallic gap in the minority-spin channel is reduced compared to the ordered $Y$ structure, however the high spin polarization at $E_\mathrm {F}$ is still maintained for all the cases. The half-metallic nature is found to be unaffected in the $B$2 structure too. This robustness of half-metallicity against various atomic disorder could make IrCrMnZ systems more suitable in possible spintronics applications. 
 Previously, there are several Co-based Heusler alloys are reported to maintain their half-metallicity against atomic disorders.\cite{umetsu-jpcm,prb-miura-2004,jap-miuara-2004,jap-miura-2006}

 Figure 5 compares the stability of formation of disorders with respect to the ordered structure. It is found that ordered configuration is the most stable one. For each case, the formation of Cr-Mn and Ir-Cr disorder ($XA$ and $L$2$_1$-I structure) from the ordered phase ($Y$ structure) costs about 200 meV/f.u. energy.
 However, the energy difference between ordered structure and that resulting from Mn-Z disorder is remarkably higher for Z = Si and Ge (389 meV/f.u., 372 meV/f.u.), compared to that of Z= Al or Ga (98 meV/f.u., 156 meV/f.u.).
 This trend is quite consistent with the previous study on Mn$_2$RuZ Heusler alloys.\cite{troy-jpcm} For all the cases formation of $B$2-structure is the energetically most expensive. 
 Note that $\mu_{total}$ of the IrCrMnZ systems do not show any visible change for various disordered cases with respect to the ordered one, owing to the robustness of half-metallic nature against the considered swapping disorders.
 
 Figure 6 shows a comparison of total DOS for each cases of IrCrMnZ (Z= Al, Ga, Si, Ge), with considering the effect of ``spin-orbit-coupling'' (SOC) and without SOC. As shown in figure 6, the DOS around $E_\mathrm {F}$ is hardly affected by SOC. Therefore, we do not include SOC in the following part of the discussion.

\subsection{Heterojunction with MgO}
\subsubsection{Electronic structure}
In this section, we mainly discuss about the electronic structure of the IrCrMnZ/MgO (001) heterojunction. We studied IrCrMnZ/MgO (001) heterojunctions for both IrCr- and MnZ-terminated interfaces, in which Ir and Cr or Mn and Z atoms are located on the top of O atoms of MgO depending on the terminations, respectively. We found that MnZ-terminated interface is energetically more favorable over the IrCr-terminated one, which is consistent with the existing literature for related heterojunctions.\cite{troy-jmmm-2020,Miura-prb} Thus, we only include the results for the MnZ-terminated interface. The lattice mismatch at the IrCrMnZ/MgO-interface are 0.9\%, 1.2\%, -1.2\%, and 0.7\% for Z=Al, Ga, Si, and Ge, respectively.

\begin{figure}[h]
\includegraphics[width=0.7\textwidth]{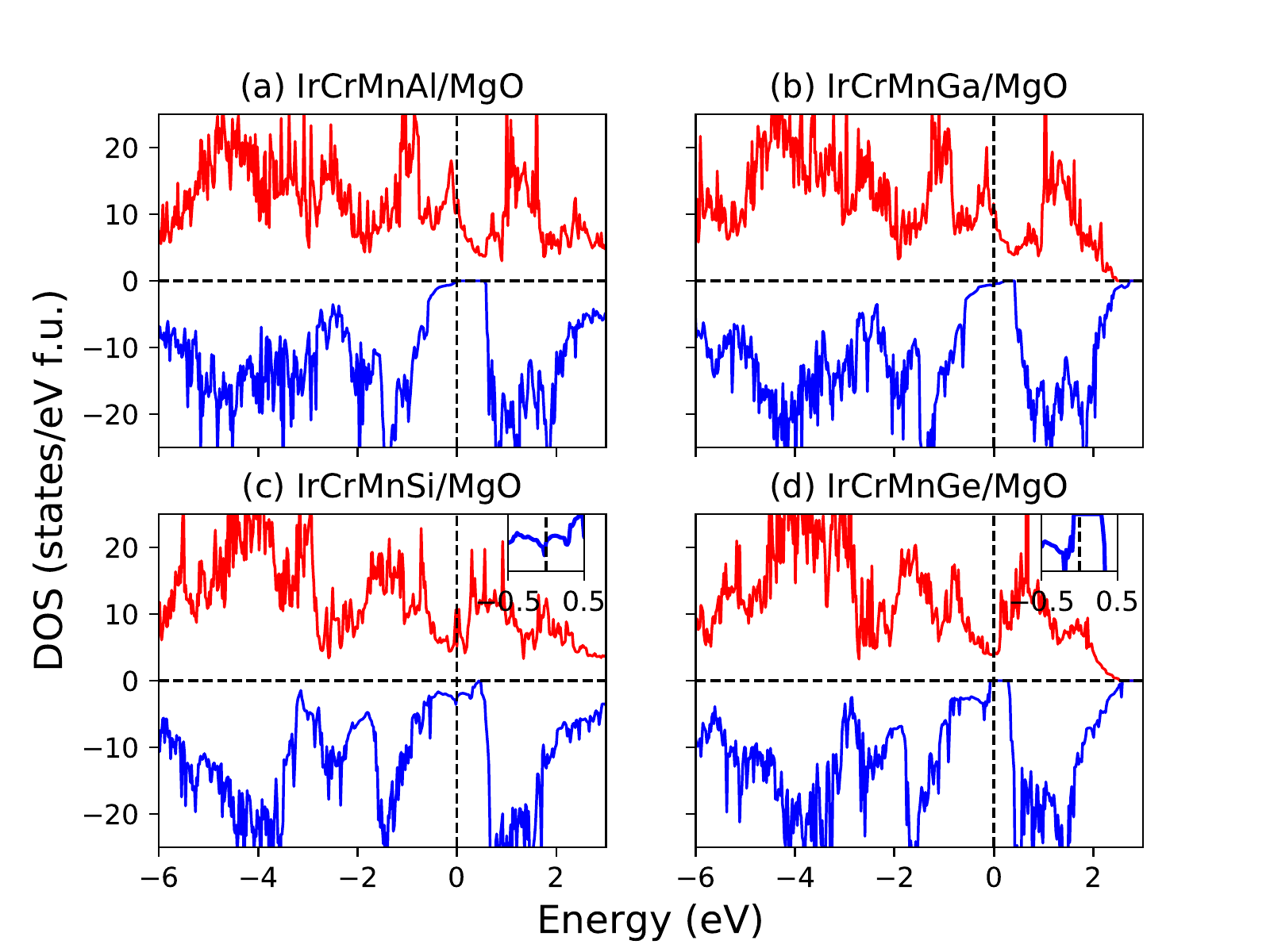}
\caption
{(Color online) The spin-resolved and total density of states (DOS) for the MnZ-terminated interface, (Z=Al, Ga, Si, Ge depending on the systems) for (a) IrCrMnAl/MgO, (b) IrCrMnGa/MgO, (c) IrCrMnSi/MgO, and (d) IrCrMnGe/MgO, respectively. The Fermi level is at 0 eV.} 

\end{figure}

Figure 7 depicts the spin-resolved DOS of the respective heterojunctions. The spin polarization at the IrCrMnAl/MgO, IrCrMnGa/MgO, IrCrMnSi/MgO, and IrCrMnGe/MgO interfaces is 99.5\%, 91\%, 36\%, and 100\%, respectively. It is evident that apart from IrCrMnSi/MgO heterojunction, in all other cases the high spin polarization of the bulk phase of these electrode materials are preserved. In the inset of figures 7 (c) and (d), we present the DOS of the minority-spin channel in the vicinity of $E_\mathrm {F}$. There is a peak at $E_\mathrm {F}$ in the DOS of in-gap states for IrCrMnSi/MgO, which reduces the spin polarization at $E_\mathrm {F}$. On the other hand, in case of IrCrMnGe/MgO, this peak is located just about 0.2 eV below $E_\mathrm {F}$, which results in 100\% spin polarization at $E_\mathrm {F}$ even in the vicinity of the interface. In both the cases of IrCrMnAl/MgO and IrCrMnGa/MgO, there are no in-gap states around $E_\mathrm {F}$ in the minority-spin channel. This difference in the interfacial electronic structure for the heterjunctions could be better understood in terms of the interfacial bond length or the characteristics of chemical bond of the respective systems, which will be discussed later.

\begin{figure}[h]
\includegraphics[width=0.7\textwidth]{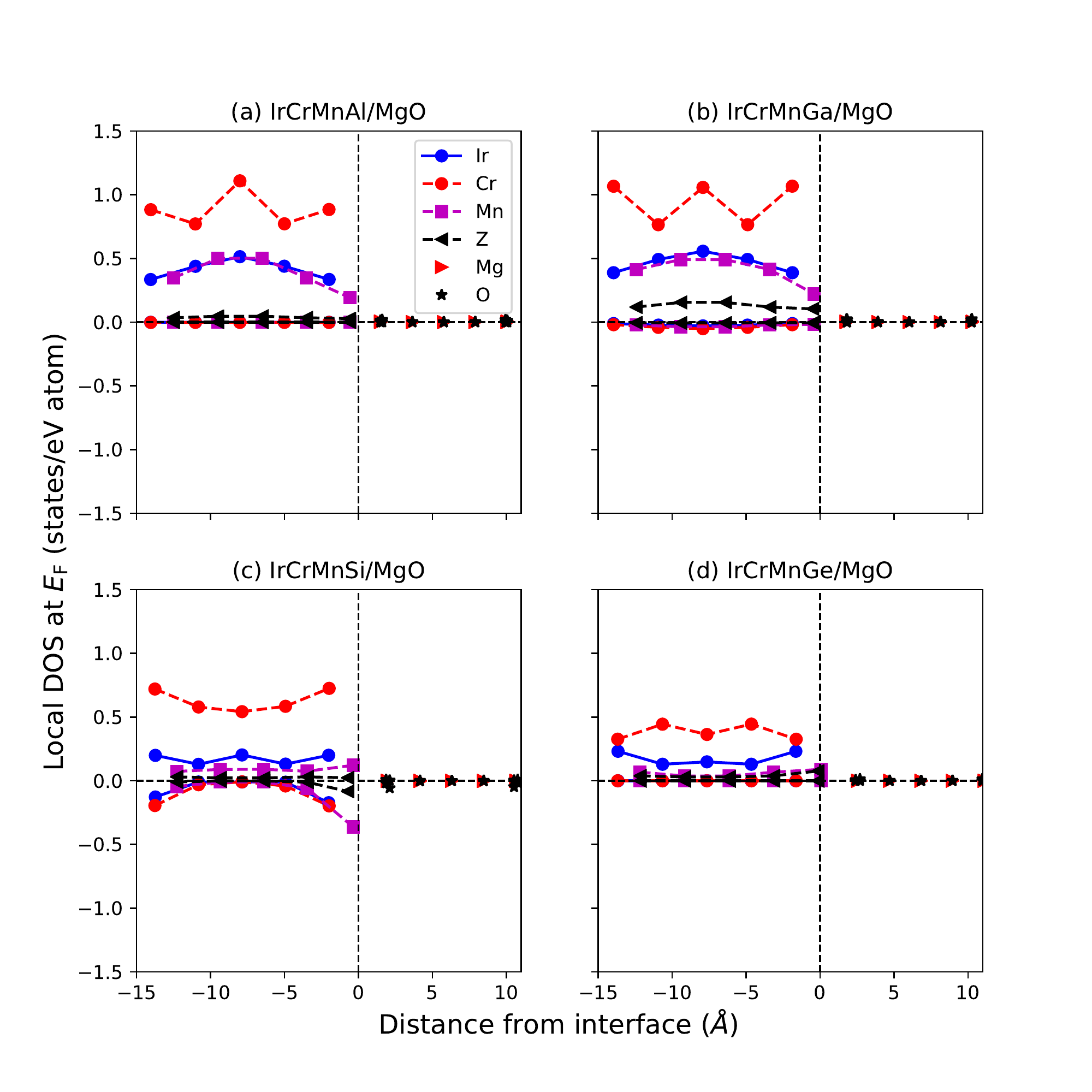}
\caption
{(Color online) Local density of states of each atom plotted as a function of distance for (a) IrCrMnAl/MgO, (b) IrCrMnGa/MgO, (c) IrCrMnSi/MgO, and (d) IrCrMnGe/MgO, respectively. } 

\end{figure}

In figure 8 we present, how spin polarization at $E_\mathrm {F}$ varies as a function of the distance from the IrCrMnZ/MgO interface. It is important to note that the interfacial states appear at $E_\mathrm {F}$ in the minority-spin gap, only for IrCrMnSi/MgO, as already mentioned. For further understanding of the interfacial states, we present the local DOS of the interfacial Mn atoms for each cases in figure 9. It is observed that in case of IrCrMnSi and IrCrMnGe there are small peaks of DOS at/around $E_\mathrm {F}$ from the interfacial Mn atoms (insets of figures 9(c) and 9(d)). Namely, the minority-spin in-gap states are predominantly originated from the interfacial Mn 3$d$ orbitals. On the other hand, IrCrMnAl and IrCrMnGa do not have such features around $E_\mathrm {F}$.

\begin{figure}[h]
\includegraphics[width=0.7\textwidth]{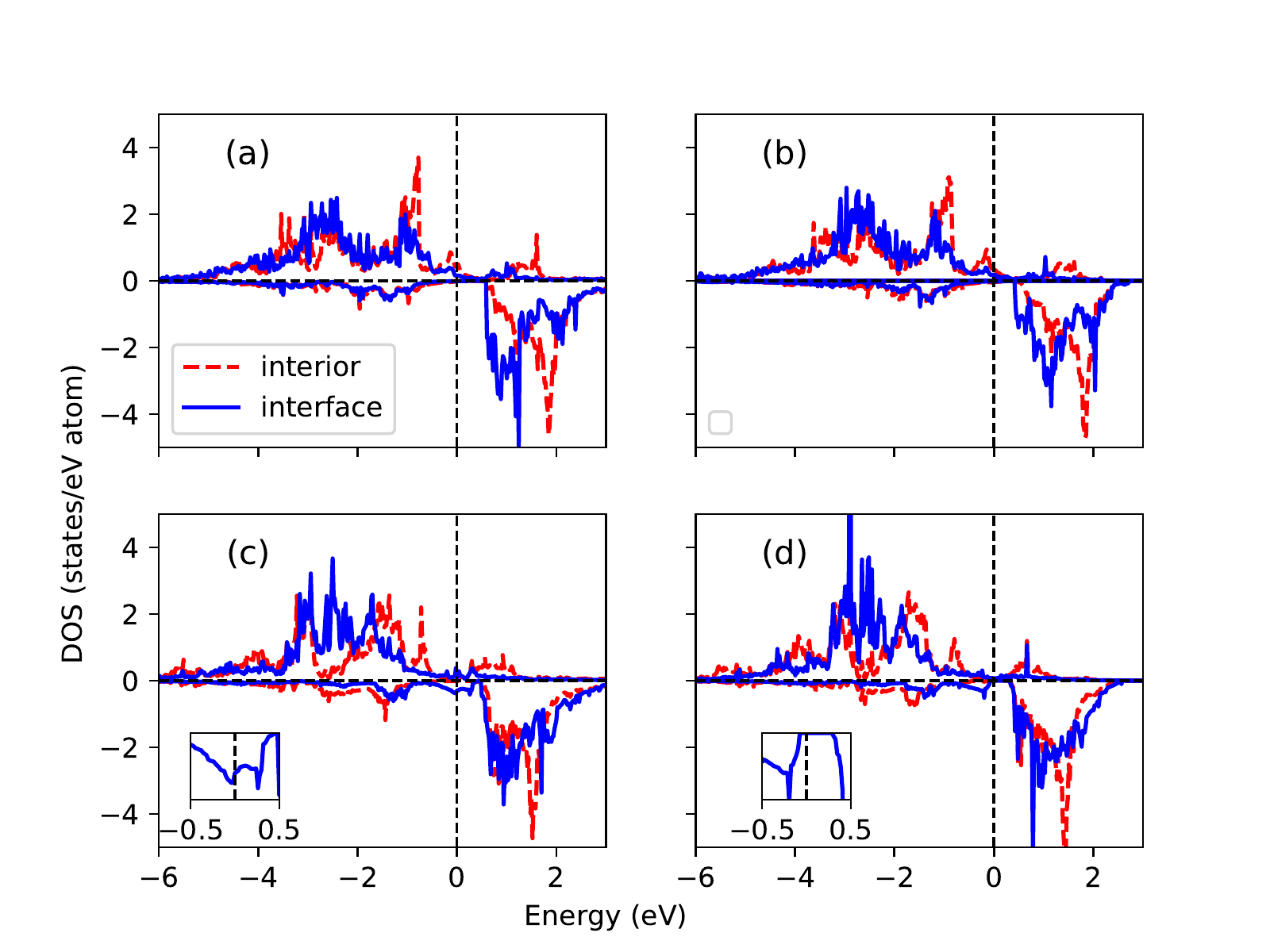}
\caption
{(Color online) Solid lines (blue) show the local density of states of interfacial Mn atoms for (a) IrCrMnAl/MgO, (b) IrCrMnGa/MgO, (c) IrCrMnSi/MgO, and (d) IrCrMnGe/MgO, respectively. In each case the local density of states of Mn atom from interior part is also presented for comparison by dashed (red) lines. The Fermi level is at 0 eV.} 

\end{figure}

\begin{table*}[hbt!]
\renewcommand{\thetable}{\arabic{table}}

\centering
\caption{Magnetic moments of Mn atom in the interfacial and interior regions of the heterojunction.}
\begin{tabular}{|c|c|c|}
\hline Material &  Interfacial Mn ($\mu_{\mathrm B}$)  &   Interior Mn ($\mu_{\mathrm B}$)\\
\hline IrCrMnAl/MgO &3.56&3.20\\
\hline IrCrMnGa/MgO &3.68&3.26 \\
\hline IrCrMnSi/MgO &3.67&2.99\\
\hline IrCrMnGe/MgO &3.80&3.22\\

\hline
\end{tabular}   
\end{table*}

We summarized the magnetic moments of the Mn atoms in the interfacial and interior regions of each heterojunction in Table 4. It is to be noted that in each case the magnetic moment of the interfacial Mn atom is larger than that of the interior region. It could be understood from figure 9, in which DOS of the Mn atoms in the interior region of heterojunction is also presented for the sake of comparison with that in the interfacial region. 
It is evident that the DOS of interfacial Mn atoms in all the cases are pushed towards the more negative energy side for the majority-spin states, leading to more localized nature of the Mn 3$d$ electrons and the enhancement of the Mn spin moment compared to the interior region. This is consistent with the results of the previous studies.\cite{Miura-prb} It is argued that at the interfacial layer there is a charge transfer from the minority-spin channel to majority-spin channel, which results in an enhancement of magnetic moment of interfacial Mn atom relative to its interior counterpart.\cite{Miura-prb}

\begin{figure}[h]
\includegraphics[width=0.7\textwidth]{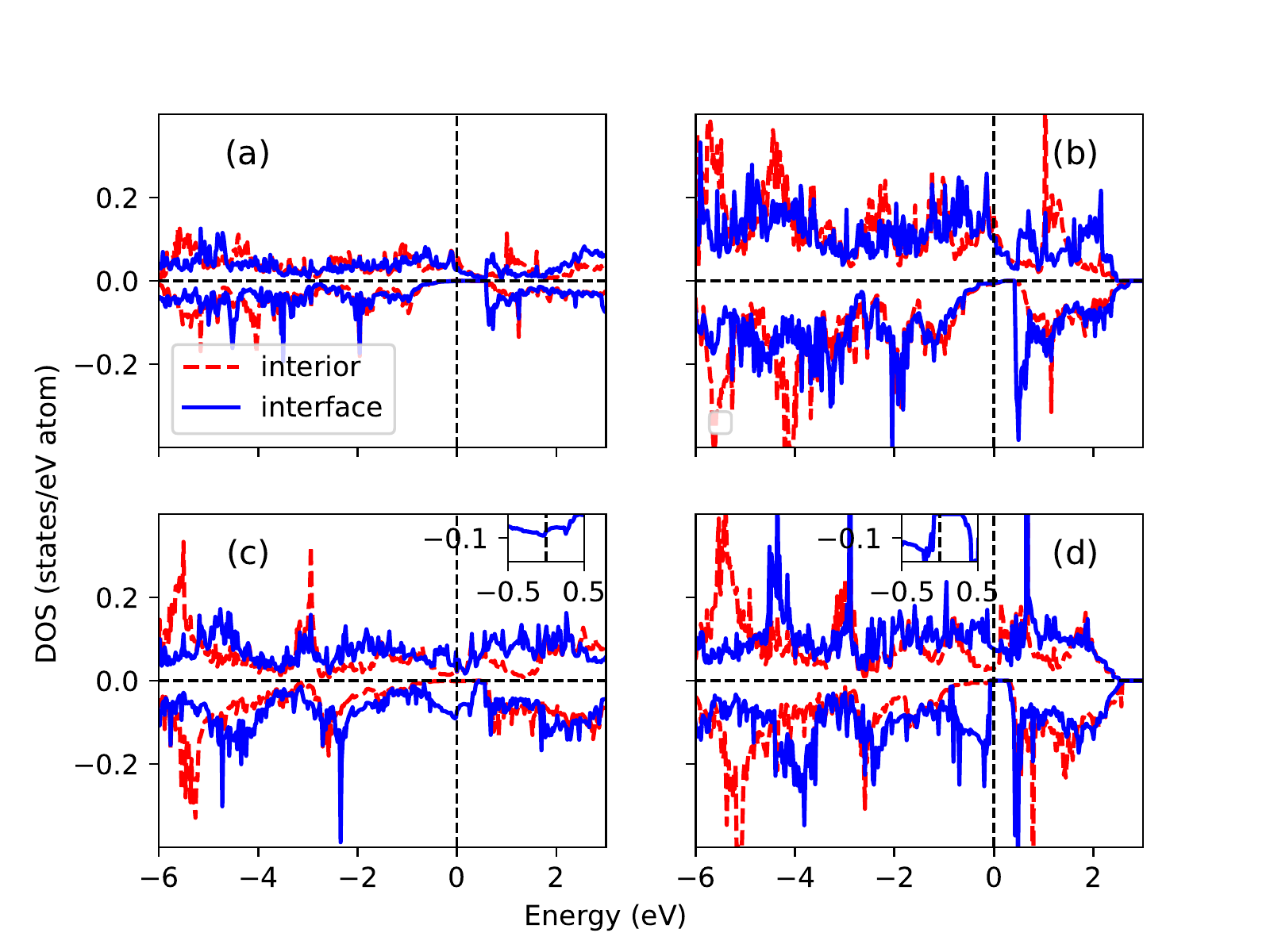}
\caption
{(Color online) Solid lines (blue) show the local density of states of interfacial Z atoms for (a) IrCrMnAl/MgO, (b) IrCrMnGa/MgO, (c) IrCrMnSi/MgO, and (d) IrCrMnGe/MgO, respectively. In each case the local density of states of Z atom from interior part is also presented for comparison by dashed (red) lines. The Fermi level is at 0 eV.} 

\end{figure}

\begin{table*}[hbt!]
\renewcommand{\thetable}{\arabic{table}}

\centering
\caption{Interfacial bond lengths.}
\begin{tabular}{|c|c|c|c|}
\hline Material &  Mn-O (\AA)  &   Z-O  (\AA)& $\delta l$ (\AA)\\
\hline IrCrMnAl/MgO &2.18&2.08& 0.10\\
\hline IrCrMnGa/MgO &2.20&2.33& 0.13 \\
\hline IrCrMnSi/MgO &2.23&2.84& 0.61\\
\hline IrCrMnGe/MgO &2.34&2.84& 0.50\\

\hline
\end{tabular}   
\end{table*}

Figure 10 shows the electronic structure of Z atoms in the interfacial and the interior regions of IrCrMnZ/MgO (001) heterojunctions. It is found that similar to the interfacial Mn atoms of IrCrMnSi/MgO and IrCrMnGe/MgO, there are interfacial states at/around $E_\mathrm {F}$ in the minority-spin gap of Si and Ge atoms. Nevertheless, the Z atoms in all the cases maintains half-metallicity in the interior region. 

It is worth investigating the physical origin of the interfacial states in the minority-spin channel from the Mn and Z atoms, for Z=Si and Ge. We tabulate the interfacial Mn-O and Z-O bond lengths for all the cases in Table 5. Note that the Mn-O bond lengths, in the cases for IrCrMnAl/MgO (2.18 \AA) and IrCrMnGa/MgO (2.20 \AA) are comparable to Al-O (2.08 \AA) and Ga-O (2.33 \AA) bond lengths. However, Si-O (2.84 \AA) and Ge-O (2.84 \AA) bond lengths are remarkably larger compared to that of Mn-O bond-lengths for IrCrMnSi/MgO (2.23 \AA) and IrCrMnGe/MgO (2.34 \AA). 
We define $\delta l$ to quantify the buckling at the MgO-interface, which is the difference of the interfacial bond lengths between Z-O and Mn-O. 
Note that, there is a significant buckling at the interface due to weak bonding between Z and O atoms for Z=Si and Ge. This weak bonding at the interface results in the interfacial states in the minority-spin gap for IrCrMnSi/MgO and IrCrMnGe/MgO. This observation is quite consistent with the interfacial electronic structure of MnGe-terminated Mn$_2$RuGe/MgO (001), and MnSi-terminated interfaces of Co$_2$MnSi/MgO (001), and CoIrMnSi/MgO (001) as discussed in the previous studies.\cite{Miura-prb,troy-jpcm,troy-jmmm-2020} From these previous reports on Heusler-alloy/MgO heterojunctions and as well as the present study, it can be inferred that interfacial structure and bonding characteristics play a crucial role in preserving the half-metallicity of electrode in its bulk phase. These studies show, in case of Heulser alloys, when Z is a group $IV$ elements (e.g. Si or Ge), there is a remarkable buckling at the MgO interface, which results in interfacial states at the minority gap. This problem is overcome for Z=Al, or Ga as observed in the cases Co$_2$CrAl/MgO,\cite{Miura-prb} CoIrMnAl/MgO,\cite{troy-jmmm-2020} IrCrMAl/MgO, and IrCrMnGa/MgO. In spite of half-metallic interface for Co$_2$CrAl/MgO and CoIrMnAl/MgO, 
 the major drawback for these materials was the absence of majority-spin $\Delta_1$ band across $E_\mathrm {F}$, which forbids coherent tunneling. In case of IrCrMnAl and IrCrMnGa, the majority-spin $\Delta_1$ band crosses $E_\mathrm {F}$ and it maintains half-metallicity at the MgO-interface, simultaneously.

\begin{figure}[h]
\includegraphics[width=0.8\textwidth]{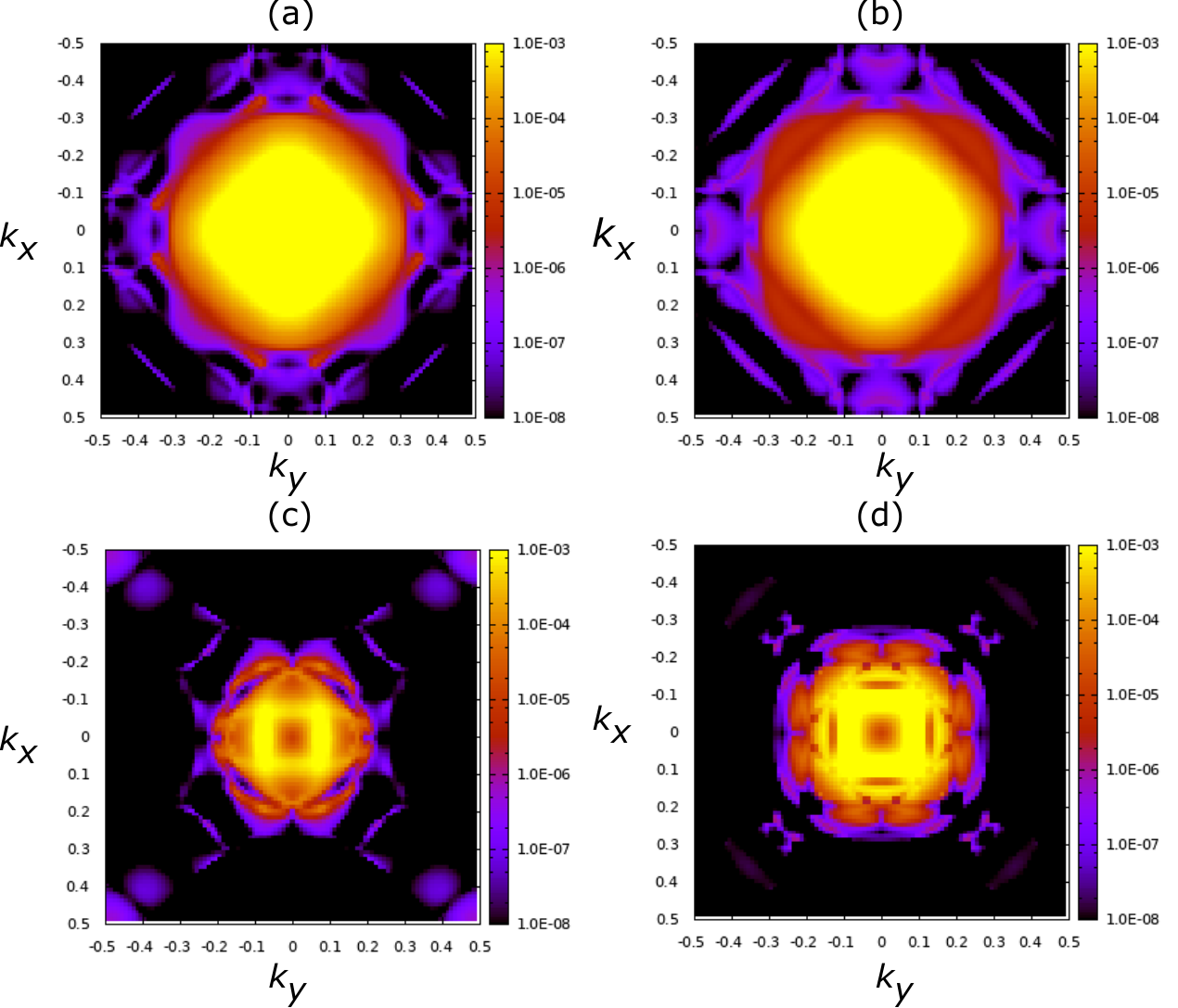}
\caption
{(Color online)In plane wave-vector $k_{\parallel}= (k_x, k_y)$ dependence of majority-spin transmittance for (a) IrCrMnAl/MgO/IrCrMnAl, (b) IrCrMnGa/MgO/IrCrMnGa, (c) IrCrMnSi/MgO/IrCrMnSi, (d) IrCrMnGe/MgO/IrCrMnGe in the parallel magnetization configuration. Here, we consider five monolayers of MgO barrier.} 
\end{figure}

\subsubsection{ Tunneling transport properties}
Finally, in figure 11 we present the majority-spin transmittance of the IrCrMnZ/MgO/IrCrMnZ (001) MTJs as a function of in-plane wavevector, $k_{\parallel}= (k_x, k_y)$, at $E_\mathrm {F}$ for parallel magnetization configuration. Here, we consider five monolayers of MgO barrier. Owing to the presence of partially occupied $\Delta_1$ band, we find there is a finite transmittance at $k_{\parallel}=0$ in each case. However, the conductance are significantly large in the cases of IrCrMnAl/MgO/IrCrMnAl, IrCrMnGa/MgO/IrCrMnGa, compared to the rest of the two cases IrCrMnSi/MgO/IrCrMnSi, and IrCrMnGe/MgO/IrCrMnGe. One of the reason behind this large parallel conductance for the first two MTJs could be the smoother interfacial structure and strong bonding across the interface with MgO, which helps coupling of the $\Delta_1$ states to insulating MgO layer more effectively, compared to the rest of the two cases. The values of parallel conductance for IrCrMnAl/MgO/IrCrMnAl, IrCrMnGa/MgO/IrCrMnGa, IrCrMnSi/MgO/IrCrMnSi, IrCrMnGe/MgO/IrCrMnGe are 1.32$\times$10$^{-3}$, 8.41$\times$10$^{-4}$, 4.35$\times$10$^{-5}$, and 1.04$\times$10$^{-4}$, respectively, where, all these values are in the units of $e^2/h$.

 \section{Conclusion}

Based on first-principles calculations, here we investigated the feasibility of IrCrMnZ (Z=Al, Ga, Si, Ge) as electrode materials of MgO-based MTJs. 
Phonon dispersion curves with no negative frequency confirm the dynamical stability of the systems. Negative formation energy and phase separation energy signify the stability of these systems against possible decomposition into other alloys/compounds. Opposite spin alignments of nearest neighboring Mn and Cr atoms ensures low magnetization of these systems. It also results in a strong antiferromagnetic exchange coupling between Mn and Cr, hence high $T_\mathrm {C}$, which are beneficial in device application. IrCrMnZ (Z=Al, Ga, Si, Ge) system have very high spin polarization.
High spin polarization is maintained even at the interface with MgO, except for IrCrMnSi/MgO heterojunction. Lattice mismatch with MgO is about 1\% for all the cases, which could be beneficial in the formation of dislocation free heterojunction, to have better spin-transport property. Most importantly, the presence of highly spin-polarized $\Delta_1$ band, 
very high $T_\mathrm {C}$ (above 1350 K), specifically for IrCrMnAl and IrCrMnGa, and large majority-spin conductance via the IrCrMnZ/MgO/IrCrMnZ MTJs for parallel magnetization configuration, suggest that IrCrMnAl and IrCrMnGa could be promising electrode materials for MgO-based MTJs.

\acknowledgments 
This work was supported in part by JST CREST (No. JPMJCR17J5) and by CSRN, Tohoku University.
 The authors are grateful to S. Mizukami, A. Hirohata, T. Tsuchiya, H. Shinya, and T. Ichinose for valuable discussion. 

\section*{Data Availability Statement} The data that supports the findings of this study are available within the article.

{}

\clearpage

\end{document}